\documentclass{ws-procs9x6}

\begin{document}

\title{The formation of spiral arms and rings in barred galaxies}

\author{M. Romero-G\'omez$^*$ and E. Athanassoula}

\address{Laboratoire d'Astrophysique de Marseille,\\ Observatoire
Astronomique de Marseille Provence,\\ 2 Place Le Verrier 13248 
Marseille, France\\
$^*$E-mail: merce.romerogomez@oamp.fr}

\author{J.J. Masdemont}

\address{I.E.E.C \& Dep. Mat. Aplicada I, Universitat Polit\`ecnica de Catalunya,\\
Av. Diagonal 647, 08028 Barcelona, Spain}

\author{C. Garc\'ia-G\'omez}

\address{D.E.I.M., Universitat Rovira i Virgili,\\ Av. Pa\"{i}sos Catalans 26, 
43007 Tarragona, Spain}

\begin{abstract}
We propose a new theory to explain the formation of spiral arms and of
all types of outer rings in barred galaxies. We have extended and applied
the technique used in celestial mechanics to compute transfer orbits. Thus,
our theory is based on the chaotic orbital motion driven by the invariant
manifolds associated to the periodic orbits around the hyperbolic equilibrium
points. In particular, spiral arms and outer rings are related to the
presence of heteroclinic or homoclinic orbits. Thus, $R_1$ rings are associated
to the presence of heteroclinic orbits, while $R_1R_2$ rings are associated 
to the presence of homoclinic orbits. Spiral arms and $R_2$ rings, however, 
appear when there exist neither heteroclinic nor homoclinic orbits. We 
examine the parameter space of three realistic, yet simple, barred galaxy 
models and discuss the formation of the different morphologies according to the
properties of the galaxy model. The different morphologies arise from
differences in the dynamical parameters of the galaxy.
\end{abstract}


\bodymatter

\section{Introduction}\label{sec:intro}
Bars are very common features in disk galaxies. According to Eskridge {\it et al.} 
[\refcite{esk00}] in the near infrared $56\%$ of the galaxies are strongly barred and 
$6\%$ are weakly barred. A large fraction of barred galaxies show either spiral 
arms emanating from the ends of the bar or spirals that end up forming
outer rings (Elmegreen \& Elmegreen [\refcite{elm82}]; Sandage \& Bedke [\refcite{san94}]).

Spiral arms are believed to be density waves (Lindblad [\refcite{lind63}]). 
Toomre [\refcite{too69}], finds that the spiral arms are density waves that propagate 
outwards towards the principal Lindblad resonances, where they damp. So other mechanisms
for replenishment are needed (see for example Lindblad [\refcite{lind60}]; Toomre 
[\refcite{too69,too81}]; Toomre \& Toomre [\refcite{too72}]; Sanders \& Huntley 
[\refcite{san76}]; Athanassoula [\refcite{ath84}] for more details). Rings have 
been studied by Schwarz [\refcite{sch81,sch84,sch85}]. The author studies the response 
of a gaseous disk galaxy to a bar-like perturbation. He relates the rings with the 
position of the principal Lindblad resonances. There are different types of outer 
rings and they can be classified according to the relative orientation of the principal 
axes of the inner and outer rings (Buta [\refcite{but95}]). If the two axes are perpendicular, 
the outer ring has an eight-shape and it is called $R_1$ ring. If they are parallel, it 
is called $R_2$ ring. There are galaxies where both types of rings are present, in which case 
the outer ring is simply called $R_1R_2$ ring. 

Our approach is from the dynamical systems point of view. We first note that both
spiral arms and (inner and outer) rings emanate from, or are linked to, the ends of 
the bar, where the unstable equilibrium points of a rotating system are located. 
We also note that, so far, no common theory for the formation of both features has
been presented. We therefore study in detail the neighbourhood of 
the unstable points and we find that spiral arms and rings are flux tubes
driven by the invariant manifolds associated to the periodic orbits around the 
unstable equilibrium points.

This paper is organised as follows. In Sec.~\ref{sec:model}, we give the characteristics
of each component of the model and the potential used to describe it. 
In Sec.~\ref{sec:dyn}, we give the equations of motion and we study the neighbourhood of
the equilibrium points. In particular, we give definitions of the Lyapunov periodic
orbits, the invariant manifolds associated to them, and of the homoclinic and heteroclinic 
orbits. In Sec.~\ref{sec:res}, we present our results and in Sec.~\ref{sec:sum}, we 
briefly summarise.

\section{Description of the model}\label{sec:model}
\begin{figure}
\begin{center}
\psfig{file=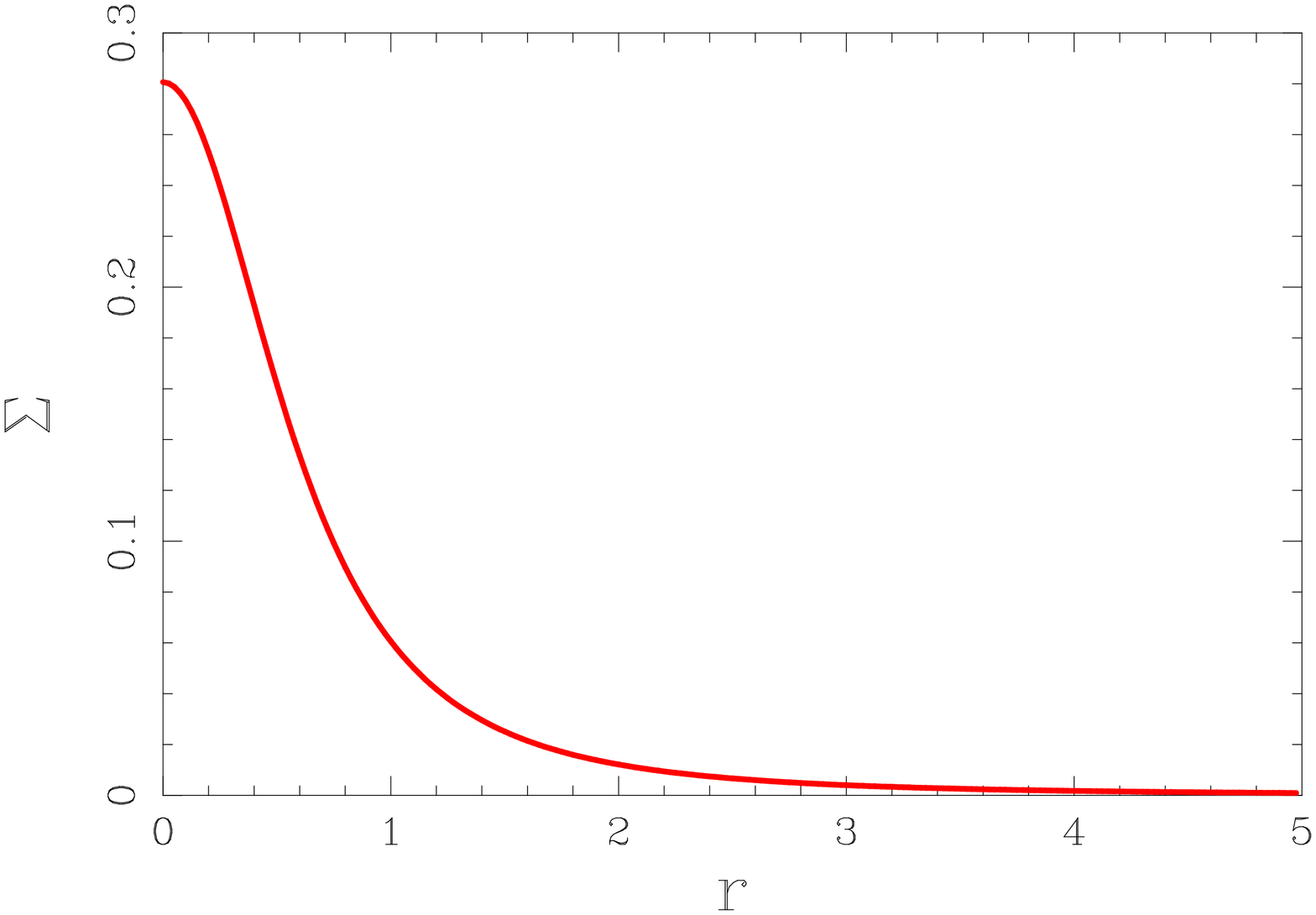,width=0.24\textwidth, angle=-90.}\hspace{0.25cm}
\psfig{file=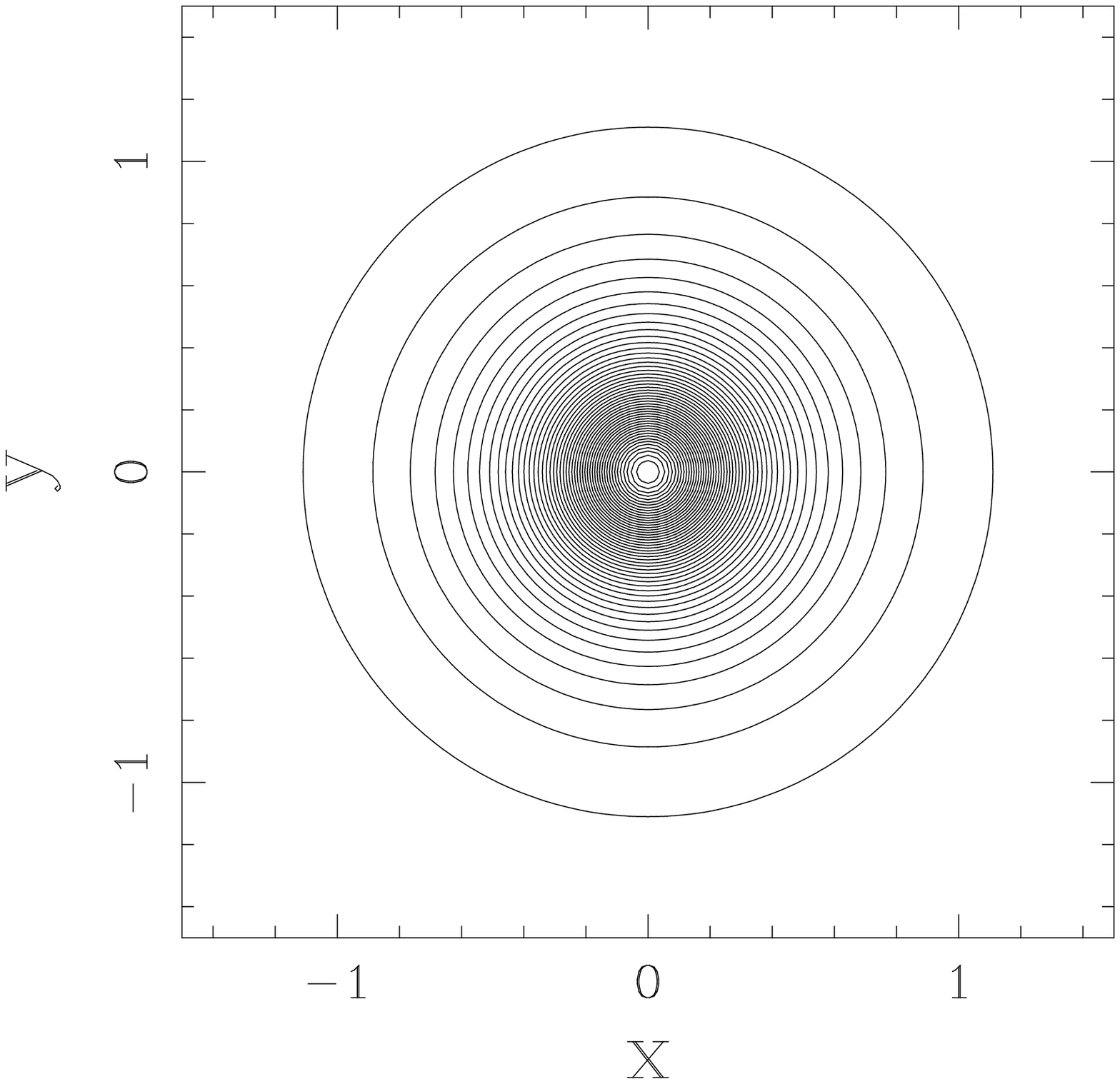,width=0.24\textwidth,angle=-90.}\hspace{0.25cm}
\psfig{file=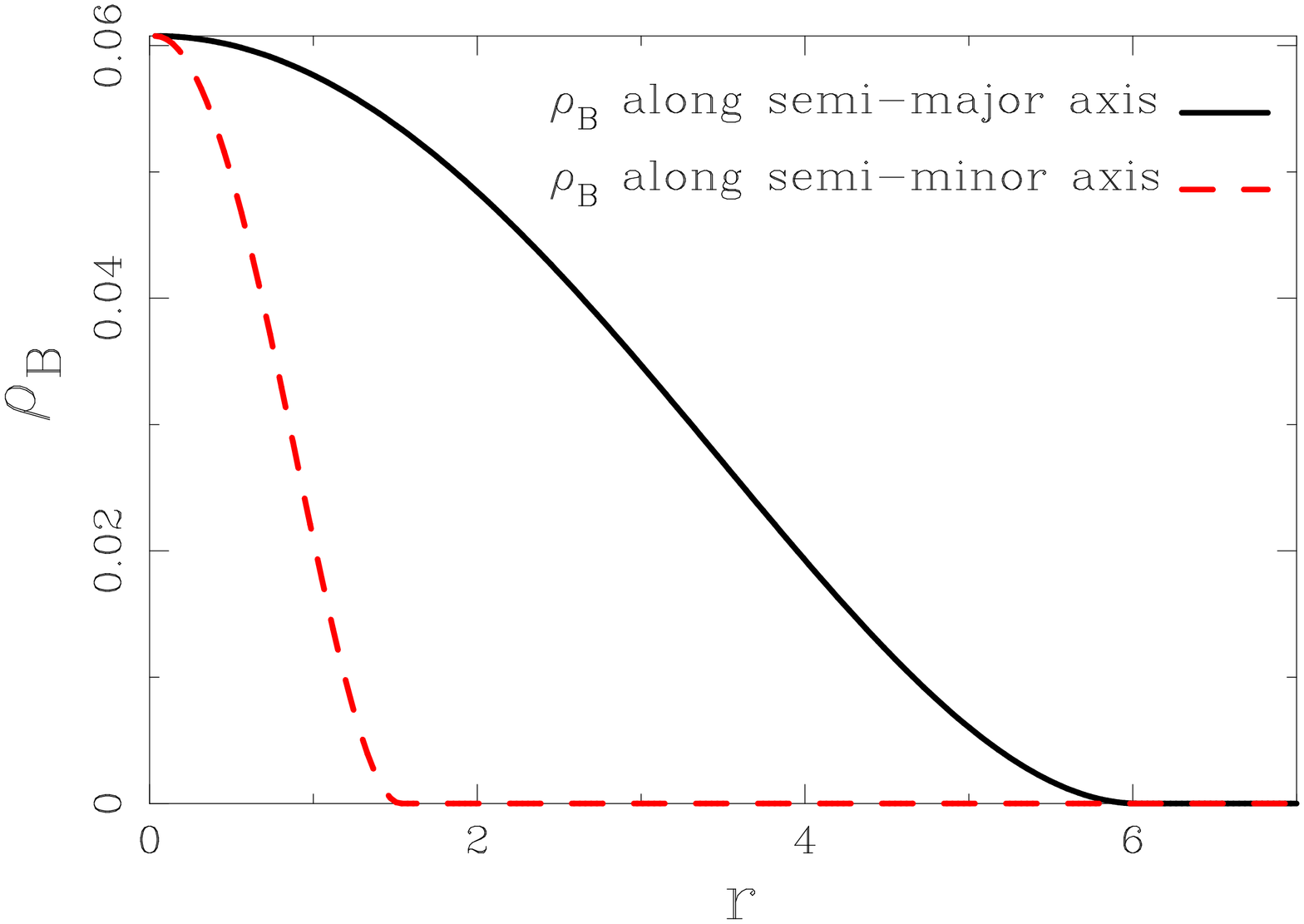,width=0.24\textwidth,angle=-90.}
\end{center}
\caption{Characteristics of the components. {\it Left panel:} Density function of 
the Kuzmin-Toomre disc (red solid line) with $r_d=0.75$ and $V_d=1.5$. {\it Middle 
panel:} Isodensity curves for the spherical distribution representing the spheroid
with parameters $r_b=0.3326$ and $\rho_b=23552.37$. {\it Right panel:} Density 
along the semi-major axis (black solid line) and the semi-minor axis 
(red dashed line) of a Ferrers bar with $n=2$, $a=6$, $b=1.5$ and $\rho_0=0.0193$.}
\label{fig:model}
\end{figure}

We use a model introduced in Athanassoula [\refcite{ath92}] that consists of the 
superposition of an axisymmetric and a bar-like component. The axisymmetric component 
is the superposition of a disc and a spheroid. The disc is modelled as a Kuzmin-Toomre 
disc (Kuzmin [\refcite{kuz56}]; Toomre [\refcite{too63}]) of surface density 
$\Sigma(r)$ (see also left panel of 
Fig.~\ref{fig:model}): 
\begin{equation}\label{eq:kuz}
\Sigma(r) = \frac{V_d^2}{2\pi r_d}\left(1+\frac{r^2}{r_d^2}\right)^{-3/2},
\end{equation}
where the parameters $V_d$ and $r_d$ set the scales of the velocities and radii of the
disc, respectively. The spheroid is modelled using a spherical density distribution, 
$\rho(r)$ (Eq.~\ref{eq:sph}), characteristic for spheroids. In the middle panel of 
Fig.\ref{fig:model}, we plot the isodensity curves for this density function:
\begin{equation}\label{eq:sph}
\rho(r)=\rho_b\left(1+\frac{r^2}{r_b^2}\right)^{-3/2},
\end{equation}
where $\rho_b$ and $r_b$ determine the central density and scale-length of the spheroid. 

Bars are non-axisymmetric features with high ellipticities. We will use three different bar
models.In the first one the bar potential is described by a Ferrers ellipsoid 
(Ferrers [\refcite{fer77}]) whose density distribution is:
\begin{equation}
\rho_{B}(x,y)=\left\{\begin{array}{lr}
\rho_0(1-m^2)^n & m\le 1\\
 0 & m\ge 1,
\end{array}\right.
\label{eq:Ferden}
\end{equation}
where $m^2=x^2/a^2+y^2/b^2$. The values of $a$ and $b$ determine the shape of
the bar, $a$ being the length of the semi-major axis, which is placed along
the $x$ coordinate axis, and $b$ being the length of the semi-minor axis. The
parameter $n$ measures the degree of concentration of the bar and $\rho_0$
represents the bar central density. In the right panel of Fig.~\ref{fig:model}, we plot the 
density function along the semi-major and semi-minor axes of the Ferrers ellipsoid
with index $n=2$, and principal axes $a=6$ and $b=1.5$. 

We also use two ad-hoc potentials, namely a Dehnen's bar type (Dehnen [\refcite{deh00}]) 
and a Barbanis-Woltjer (BW) bar type (Barbanis \& Woltjer [\refcite{bar67}]) to compare 
to the results obtained with the Ferrers ellipsoid. The Dehnen's bar potential has 
the following expression:
\begin{equation}
\Phi_1(r,\theta)=-\frac{1}{2}\epsilon v_0^2\cos(2\theta)\left\{{\begin{array}{ll}
\displaystyle 2-\left(\frac{r}{\alpha}\right)^n, & r\le \alpha\rule[-.5cm]{0cm}{1.cm}\\
\displaystyle \left(\frac{\alpha}{r}\right)^n, & r\ge \alpha,\rule[-.5cm]{0cm}{1.cm}
\end{array}}\right. 
\label{eq:phi1}
\end{equation}
where the parameter $\alpha$ is a characteristic length scale and $v_0$ is a characteristic 
circular velocity. The parameter $\epsilon$ is related to the bar strength. The BW potential 
has the expression:
\begin{equation}
\Phi_2(r,\theta)=\hat{\epsilon}\sqrt{r}(r_1-r)\cos(2\theta),
\label{eq:phi2}
\end{equation}
where the parameter $r_1$ is a characteristic scale length and $\hat{\epsilon}$ is related to 
the bar strength.

The bar-like component rotates anti-clockwise with angular velocity 
${\bf \Omega_p}=\Omega_p{\bf z}$, where $\Omega_p$ is a constant pattern
speed\, \footnote{Bold letters denote vector notation. The vector {\bf z} is a unit
vector.}.

\section{Equations of motion and dynamics around $L_1$ and $L_2$}\label{sec:dyn}
The equations of motion in a frame rotating with angular speed ${\bf \Omega_p}$ in 
vector form are
\begin{equation}\label{eq-motvec}
{\bf
\ddot{r}=-\nabla \Phi} -2{\bf (\Omega_p \times \dot{r})-  \Omega_p \times
(\Omega_p\times r)},
\end{equation}
where the terms $-2 {\bf \Omega_p\times \dot{r}}$ and $-{\bf \Omega_p \times
(\Omega_p\times r)}$ represent the Coriolis and the centrifugal
forces, respectively, $\Phi$ is the potential and ${\bf r}$ is the position vector. 
We define an effective potential 
$\Phi_{\hbox{\scriptsize eff}}=\Phi-\frac{1}{2}\Omega_p^2\,
(x^2+y^2),$ then Eq. (\ref{eq-motvec}) becomes 
${\bf \ddot{r}=-\nabla \Phi_{\hbox{\scriptsize eff}}} -2{\bf (\Omega_p \times \dot{r})},$ 
and the Jacobi constant is
\begin{equation}\label{eq-energy}
E_J = \frac{1}{2} {\bf\mid \dot{r}\mid} ^2 + \Phi_{\hbox{\scriptsize eff}},
\end{equation}
which, being constant in time, can be considered as the energy in the
rotating frame. The surface $\Phi_{\hbox{\scriptsize eff}}=E_J$ ($E_J$ defined as in 
Eq. (\ref{eq-energy})) is called the zero velocity surface, and its intersection with 
the $z=0$ plane gives the zero velocity curve. All regions in which 
$\Phi_{\hbox{\scriptsize eff}}>E_J$ are forbidden to a star with this energy, and are 
thus called forbidden regions.

\begin{figure}
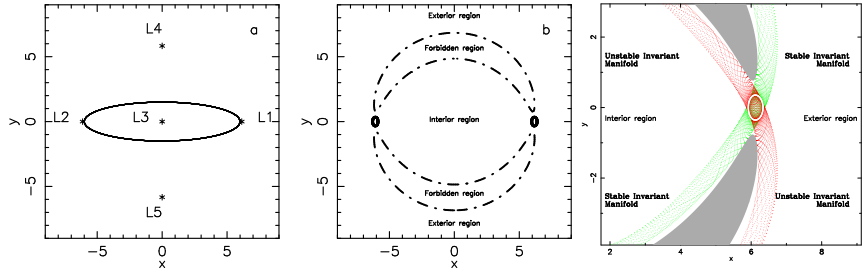

\begin{center}
\psfig{file=romerogomezfig4.ps,width=0.31\textwidth, angle=-90.}
\psfig{file=romerogomezfig5.ps,width=0.3\textwidth,angle=-90.}
\end{center}
\caption{Dynamics around the $L_1$ and $L_2$ equilibrium points. {\it Left panel:} 
Position of the equilibrium points and outline of the bar. {\it Middle panel:} 
Zero velocity curves and Lyapunov periodic orbits around $L_1$ and $L_2$. {\it Right
panel:} Unstable (in red) and stable (in green) invariant manifolds associated to 
the periodic orbit around $L_1$. In grey, we plot the forbidden region. From 
Romero-G\'omez {\it et al.} 2006, Astronomy \& Astrophysics, 453, 39, EDP Sciences.}
\label{fig:dynamics}
\end{figure}

For our calculations we place ourselves in a frame of reference corotating with the bar, 
and the bar semi-major axis is located along the $x$ axis. In this rotating frame we 
have five equilibrium points, which, due to the similarity with the Restricted Three
Body Problem, are also called Lagrangian points (see left panel of 
Fig.~\ref{fig:dynamics}). The points located symmetrically along the $x$ axis, namely 
$L_1$ and $L_2$, are linearly unstable. The ones located on the origin of coordinates, 
namely $L_3$, and along the $y$ axis, namely $L_4$ and $L_5$, are linearly stable.  
The zero velocity curve defines two different regions, namely, an exterior region 
and an interior one that contains the bar. The interior and exterior regions are 
connected via the equilibrium points (see middle panel of Fig.~\ref{fig:dynamics}). Around 
the equilibrium points there exist families of periodic orbits, e.g. around the central 
equilibrium point the well-known $x_1$ family of periodic orbits that is 
responsible for the bar structure. 

The dynamics around the unstable equilibrium points is described in detail in
Romero-G\'omez {\it et al.} [\refcite{rom06}]; here we give only a brief summary.
Around each unstable equilibrium point there exists a family of periodic 
orbits, known as the family of Lyapunov orbits (Lyapunov [\refcite{lya49}]). For a 
given energy level, two stable and two unstable sets of asymptotic orbits emanate 
from the corresponding periodic orbit, and they are known as the stable and the unstable 
invariant manifolds, respectively. The stable invariant manifold is the set of 
orbits that tends to the periodic orbit asymptotically. In the same way, the unstable 
invariant manifold is the set of orbits that departs asymptotically from the 
periodic orbit (i.e. orbits that tend to the Lyapunov orbits when the time 
tends to minus infinity), as seen in the right panel of Fig.~\ref{fig:dynamics}. Since the 
invariant manifolds extend well beyond the neighbourhood of the equilibrium 
points, they can be responsible for global structures. 

In Romero-G\'omez {\it et al.} [\refcite{rom07}], we give a detailed description
of the role invariant manifolds play in global structures and, in particular, in
the transfer of matter. Simply speaking, the transfer of matter is characterised 
by the presence of homoclinic, heteroclinic, and transit orbits.

\begin{figure}
\begin{center}
\psfig{file=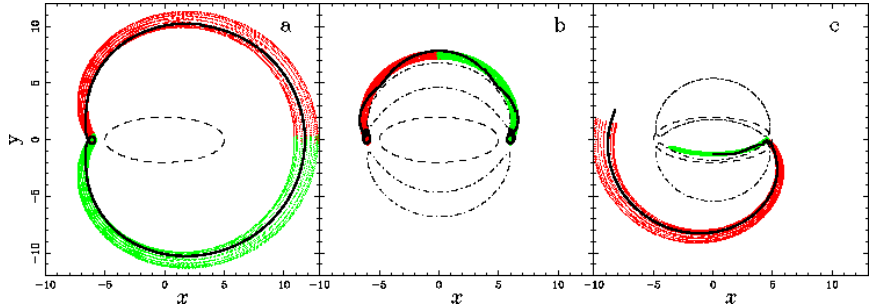,width=0.35\textwidth,angle=-90.}
\end{center}
\caption{Homoclinic {\bf (a)}, heteroclinic {\bf (b)} and transit
{\bf (c)} orbits (black thick lines) in the configuration space. In
red lines, we plot the unstable invariant manifolds associated to the
periodic orbits, while in green we plot the corresponding stable invariant
manifolds. In dashed lines, we give the outline of the bar and, in
{\bf (b)} and {\bf (c)}, we plot the zero velocity curves in dot-dashed lines.
From Romero-G\'omez {\it et al.} 2007, Astronomy and Astrophysics, 472, 63, EDP 
Sciences.}
\label{fig:transfer}
\end{figure}

Homoclinic orbits correspond to asymptotic trajectories that depart from the unstable 
Lyapunov periodic orbit $\gamma$ around $L_i$ and return asymptotically to it 
(see Fig.~\ref{fig:transfer}a). Heteroclinic orbits are asymptotic trajectories that 
depart from the periodic orbit $\gamma$ around $L_i$ and asymptotically approach the 
corresponding Lyapunov periodic orbit with the same energy around the Lagrangian point 
at the opposite end of the bar $L_j$, $i\ne j$ (see Fig.~\ref{fig:transfer}b). There also 
exist trajectories that spiral out from the region of the unstable periodic orbit, and
we refer to them as transit orbits (see Fig.~\ref{fig:transfer}c). These three types 
of orbits are chaotic orbits since they fill part of the chaotic sea when we
plot the Poincar\'e surface of section (e.g. the section $(x,\dot x)$ near $L_1$).

\section{Results}\label{sec:res}
Here we describe the main results obtained when we vary the parameters of the models
introduced in Sec.~\ref{sec:model}. One of our goals is to check separately the influence
of each of the main free parameters. In order to do so, we make families of models 
in which only one of the free parameters is varied, while the others are kept fixed. 
Our results show that only the bar pattern speed and the bar strength have an influence 
on the shape of the invariant manifolds, and thus, on the morphology of the galaxy 
(Romero-G\'omez {\it et al.} [\refcite{rom07}]). 

\begin{figure}
\begin{center}
\psfig{file=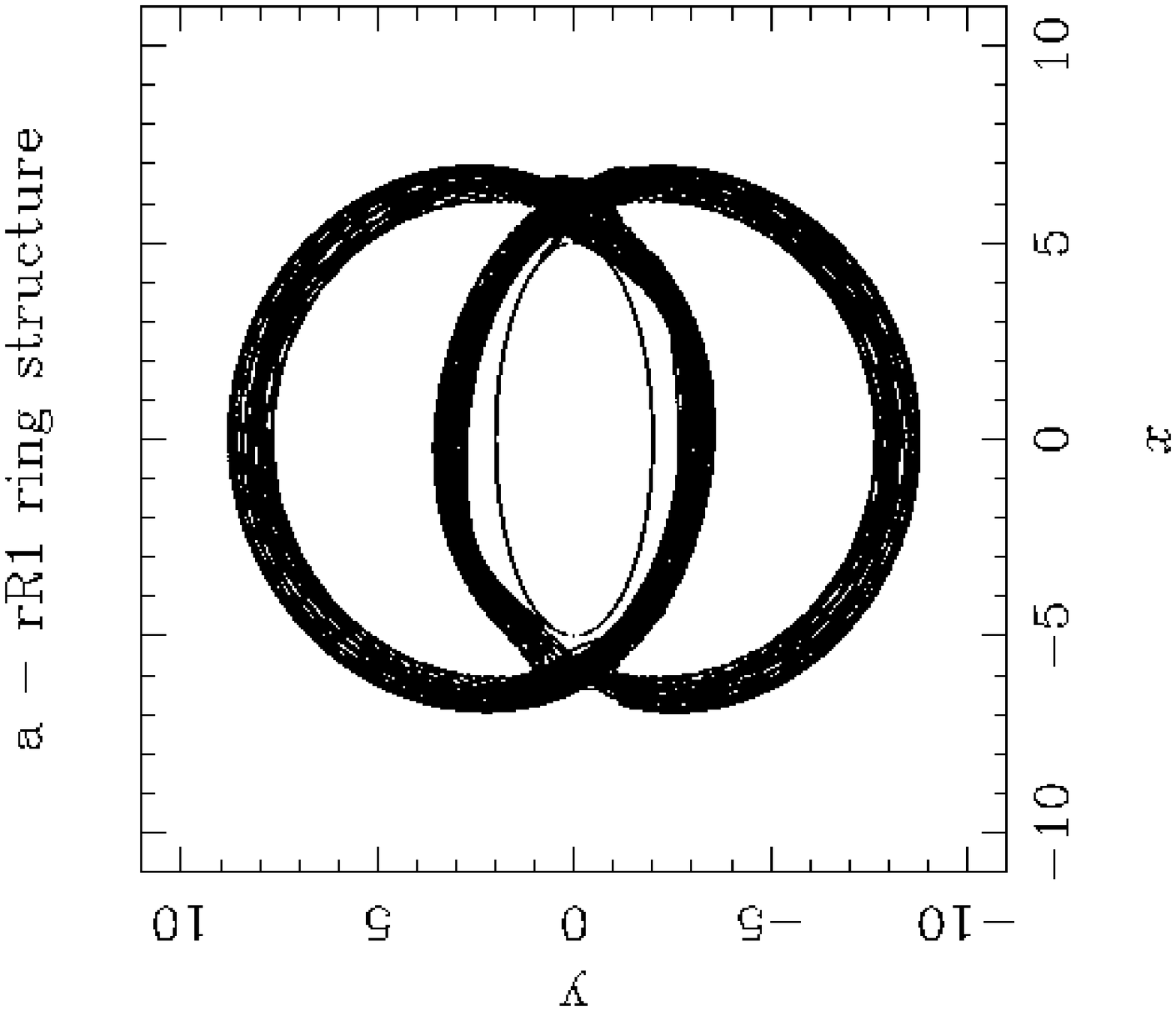,angle=-90.,width=0.23\textwidth}
\psfig{file=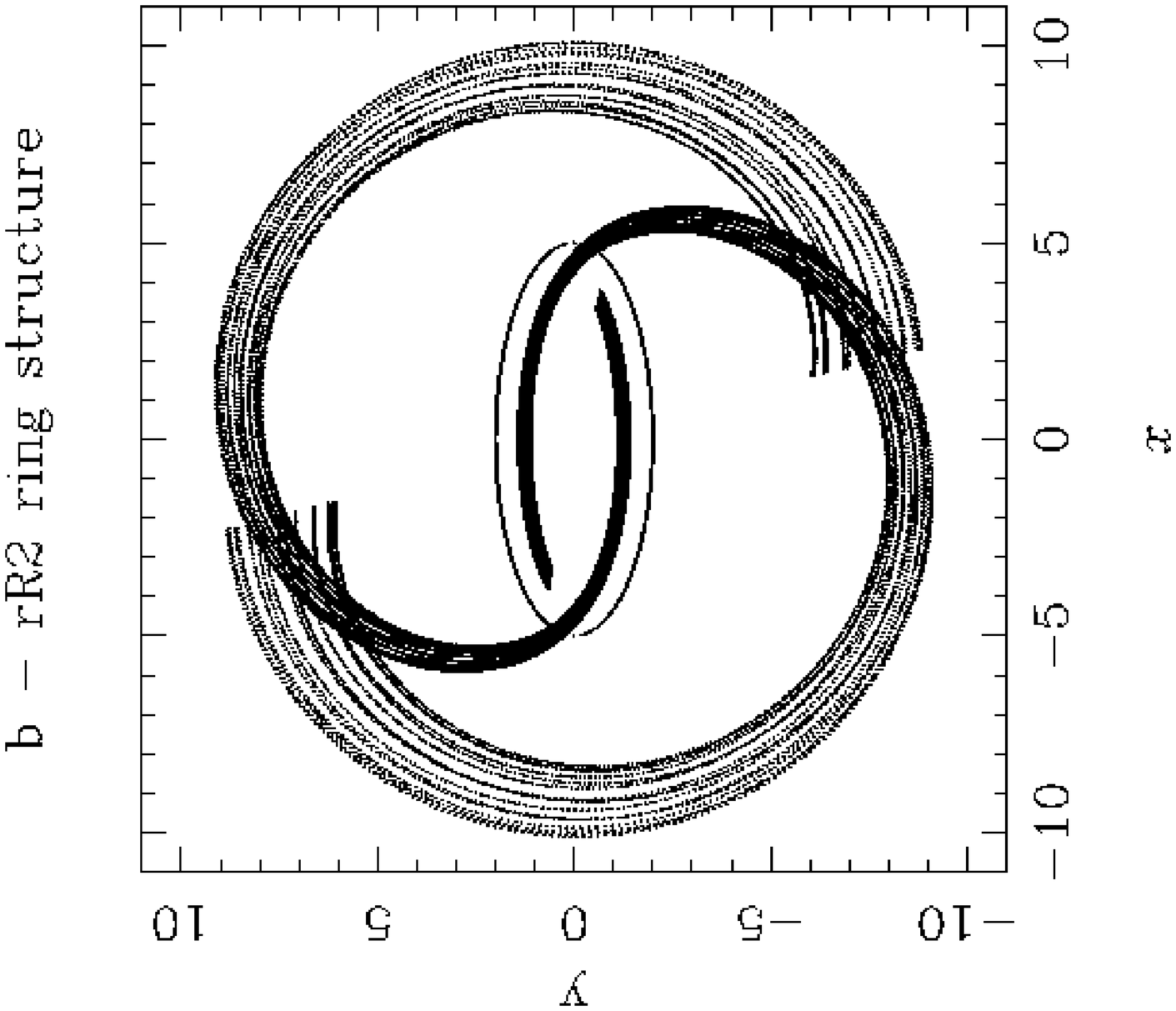,angle=-90.,width=0.23\textwidth}
\psfig{file=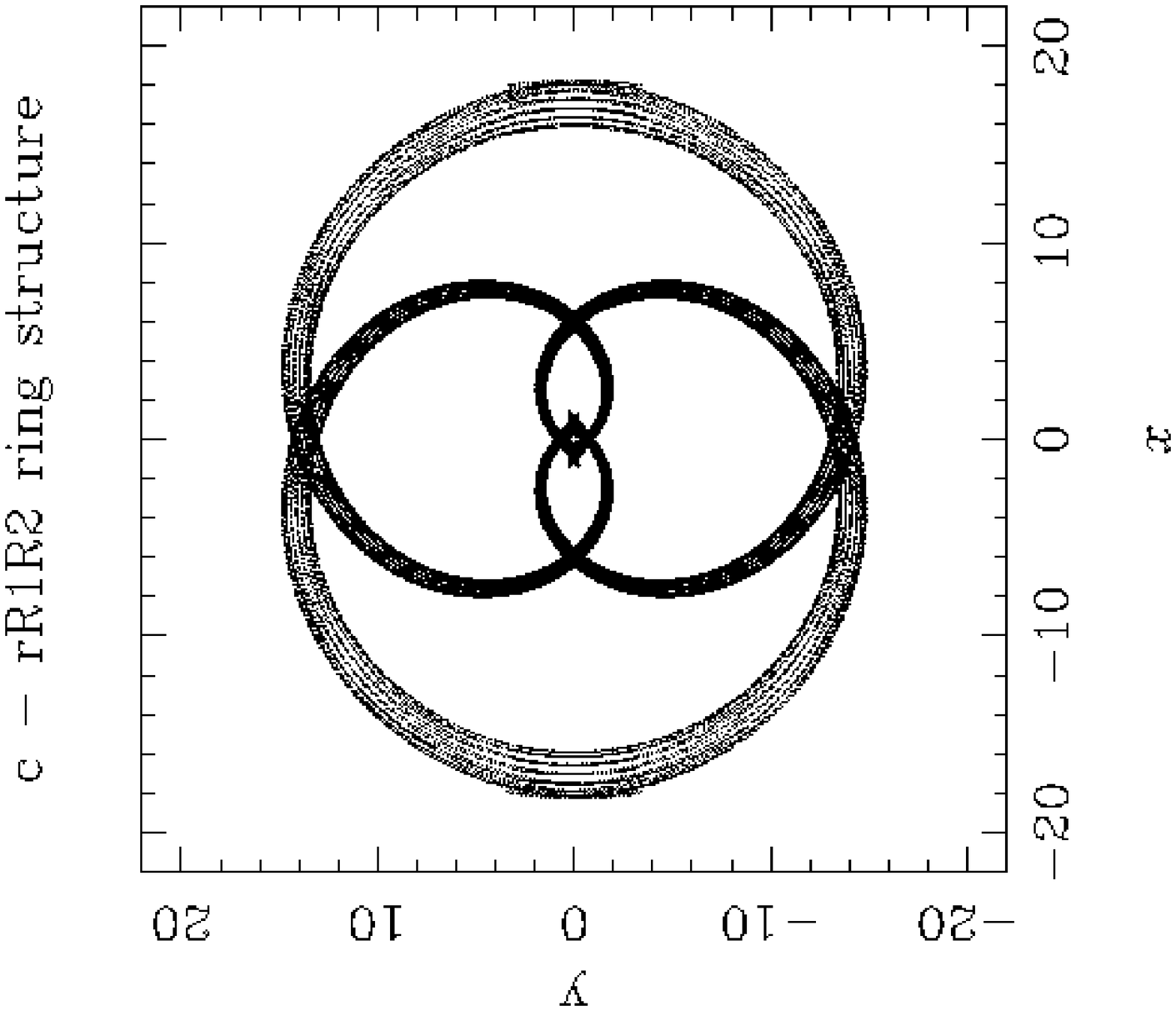,angle=-90.,width=0.23\textwidth}
\psfig{file=romerogomezfig10.ps,angle=-90.,width=0.23\textwidth}
\end{center}
\caption{Rings and spiral arms structures. We plot the invariant manifolds for different
models. {\bf (a)} $rR_1$ ring structure. {\bf (b)} $rR_2$ ring structure. {\bf (c)} 
$R_1R_2$ ring structure. {\bf (d)} Barred spiral galaxy. From Romero-G\'omez 
{\it et al.} 2007, Astronomy and Astrophysics, 472, 63, EDP Sciences.}
\label{fig:results}
\end{figure}

Our results also show that the morphologies obtained do not depend on the type of bar 
potential we use, but on the presence of homoclinic or heteroclinic orbits.
If heteroclinic orbits exist, then the ring of the galaxy is classified as $rR_1$ 
(see Fig.~\ref{fig:results}a). The inner branches of the invariant manifolds 
associated to $\gamma_1$ and $\gamma_2$ outline an inner ring that encircles the bar
and is elongated along it. The outer branches of the same invariant manifolds form an outer 
ring whose principal axis is perpendicular to the bar major axis.
If the model does not have either heteroclinic or homoclinic orbits and
only transit orbits are present, the barred galaxy will present two spiral arms
emanating from the ends of the bar. The outer branches of the unstable invariant 
manifolds will spiral out from the ends of the bar and they will not return to
its vicinity (see Fig.~\ref{fig:results}d). If the outer branches of the unstable
invariant manifolds intersect in configuration space with each other, then they form 
the characteristic shape of $R_2$ rings (see Fig.~\ref{fig:results}b). That is, the 
trajectories outline an outer ring whose principal axis is parallel to the bar major 
axis. The last possibility is if only homoclinic orbits exist. In this case, the inner 
branches of the invariant manifolds for an inner ring, while the outer branches 
outline both types of outer rings, thus the barred galaxy presents an $R_1R_2$ ring 
morphology (see Fig.~\ref{fig:results}c).

\section{Summary}\label{sec:sum}
To summarise, our results show that invariant manifolds describe well the loci 
of the different types of rings and spiral arms. They are formed by a bundle of
trajectories linked to the unstable regions around the $L_1/L_2$ equilibrium
points. The study of the influence of one model parameter on the shape of
the invariant manifolds in the outer parts of the galaxy reveals that only the pattern
speed and the bar strength affect the galaxy morphology. The study also shows
that all the different ring types and spirals can be obtained when we vary the
model parameters.

We have compared our results with some observational data. Regarding the
photometry, the density profiles across radial cuts in rings and spiral
arms agree with the ones obtained from observations. The velocities along
the ring also show that these are only a small perturbation of the circular
velocity.

\section*{Acknowledgements}
MRG acknowledges a ``Becario MAE-AECI''.


\begin{thebibliography}{9}
\bibitem{esk00} P.B. Eskridge, J.A. Frogel, R.W. Podge, A.C. Quillen, 
R.L. Davies, D.L. DePoy, M.L. Houdashelt, L.E. Kuchinski, S.V. Ram\'{\i}rez, 
K. Sellgren, D.M. Terndrup, G.P. Tiede,  {\em AJ}, {\bf 119}, 536 (2000)
\bibitem{elm82} D.M. Elmegreen, B.G. Elmegreen, {\em MNRAS}, {\bf 201}, 1021 (1982)
\bibitem{san94} A. Sandage, J. Bedke, ``The Carnegie Atlas of Galaxies'', Carnegie Inst.
Washington (1994)
\bibitem{lind63} B. Lindblad, {\em Stockholms Observatorium Ann.}, {\bf Vol. 22}, No. 5 (1963)
\bibitem{too69} A. Toomre, {\em ApJ}, {\bf 158}, 899 (1969)
\bibitem{lind60} P.O. Lindblad, {\em Stockholms Observatorium Ann.}, {\bf Vol. 21}, No. 4 (1960)
\bibitem{too72} A. Toomre, J. Toomre, {\em ApJ}, {\bf 178}, 623 (1972)
\bibitem{san76} R.H. Sanders, J.M. Huntley, {\em ApJ}, {\bf 209}, 53 (1976)
\bibitem{too81} A. Toomre, ``The structure and evolution
of normal galaxies'', eds. S.M. Fall and D. Lynden-Ball, Proc. of the Advanced Study Institute,
Cambridge, pp. 111-136 (1981)
\bibitem{ath84} E. Athanassoula, {\em Phys. Rep.}, {\bf 114}, 319 (1984)
\bibitem{sch81} M.P. Schwarz, {\em ApJ}, {\bf 247}, 77 (1981)
\bibitem{sch84} M.P. Schwarz, {\em MNRAS}, {\bf 209}, 93 (1984)
\bibitem{sch85} M.P. Schwarz, {\em MNRAS}, {\bf 212}, 677 (1985)
\bibitem{but95} R. Buta, {\em ApJS}, {\bf 96}, 39 (1995)
\bibitem{ath92} E. Athanassoula, {\em MNRAS}, {\bf 259}, 328 (1992)
\bibitem{kuz56} G. Kuzmin, {\em Astron. Zh.}, {\bf 33}, 27 (1956)
\bibitem{too63} A. Toomre, {\em ApJS}, {\bf 138}, 385 (1963)
\bibitem{fer77} N.M. Ferrers, {\em Q.J. Pure Appl. Math.}, {\bf 14}, 1 (1877)
\bibitem{deh00} W. Dehnen, {\em AJ}, {\bf 119}, 800 (2000)
\bibitem{bar67} B. Barbanis, L. Woltjer, {\em ApJ}, {\bf 150}, 461 (1967)
\bibitem{rom06} M. Romero-G\'omez, J.J. Masdemont, E. Athanassoula, C. Garc\'{i}a-G\'omez,
{\em A\& A}, {\bf 453}, 39 (2006)
\bibitem{lya49} A. Lyapunov, {\em Ann. Math. Studies}, {\bf 17} (1949)
\bibitem{rom07} M. Romero-G\'omez, E. Athanassoula, J.J. Masdemont, C. Garc\'{i}a-G\'omez, 
{\em A\& A}, {\bf 472}, 63 (2007)
\end{thebibliography}
\end{document}